\documentstyle[a4p,12pt,epsfig]{article}

\renewcommand{\Huge}{\huge}
\newcommand{\smcap}[1] {\caption[]{\small \textit{#1}}}
\newcommand{\etal}     {{\rm et al.}}

\newcommand {\costs}         {\cos\theta^*}

\newcommand{\uds} {\mathrm{uds}              }
\newcommand{\qc}  {\mathrm{c}                }
\newcommand{\qb}  {\mathrm{b}                }

\newcommand{\bb}{\mathrm{b}\overline{\mathrm{b}}}
\newcommand{\cc}{\mathrm{c}\overline{\mathrm{c}}}
\newcommand{\ee}{\mathrm{e}^+ \mathrm{e}^-}

\newcommand{\Z}{\mathrm{Z}}

\newcommand{\Dsprimepm} {\mathrm{D}^{*\prime\pm}}
\newcommand{\Dsprimep}  {\mathrm{D}^{*\prime +}}
\newcommand{\Dsprime}   {\mathrm{D}^{*\prime}}

\newcommand{\Dprime}    {\mathrm{D}^{\prime}}

\newcommand{\Dstwozero} {\mathrm{D}^{*0}_{2}}
\newcommand{\Dstwo}     {\mathrm{D}^{*}_{2}}
\newcommand{\Dzeroone}  {\mathrm{D}^{0}_{1} }
\newcommand{\Done}      {\mathrm{D}_{1} }
\newcommand{\Dpzeroone} {\mathrm{D}^{* 0}_{1}}
\newcommand{\Dsszero}   {\mathrm{D}^{(*)0}_{\mathit{J}}}
\newcommand{\Dss}       {\mathrm{D}^{(*)}_{\mathit{J}}}

\newcommand{\Dstarpm}   {\mathrm{D}^{*\pm}  }
\newcommand{\Dstarp}    {\mathrm{D}^{*+}    }
\newcommand{\Dstar}     {\mathrm{D}^{*}     }

\newcommand{\Dzero}     {\mathrm{D}^{0}     }
\newcommand{\D}         {\mathrm{D}         }

\newcommand{\B}         {\mathrm{B}         }

\newcommand{\rmrho}     {\mathrm{\rho}      }
\newcommand{\pionp}     {\mathrm{\pi}^{+}   }
\newcommand{\pionm}     {\mathrm{\pi}^{-}   }

\newcommand{\K}         {\mathrm{K}         }
\newcommand{\Km}        {\mathrm{K}^{-}     }

\newcommand{\GeVc}{\;\mathrm{GeV}/\mathit{c}}
\newcommand{\GeVcc}{\;\mathrm{GeV}/\mathit{c}^{\mathrm{2}}}
\newcommand{\MeVcc}{\;\mathrm{MeV}/\mathit{c}^{\mathrm{2}}}
\newcommand{\cm}{\;\mathrm{cm} }
\newcommand{\mm}{\;\mathrm{mm} }
\newcommand{\stddev}{\mathrm{\sigma}}

\newcommand {\xE}    {\mathit{x}_{\mathrm{E}}}
\newcommand {\dEdx}  {\mathrm{d}\mathit{E}/\mathrm{d}\mathit{x}}
\newcommand {\mDzero}{\mathit{m}_{\Dzero}}
\newcommand {\mDstar}{\mathit{m}_{\Dstar}}
\newcommand {\mDsp}  {\mathit{m}_{\Dsprime}}
\newcommand {\dm}    {\mathrm{\Delta}\mathit{m}}
\newcommand {\dmp}   {\mathrm{\Delta}\mathit{m}^{'}}

\newcommand{\GGbb}{\Gamma_{\bb}/\Gamma_{\mathrm{had}}}
\newcommand{\GGcc}{\Gamma_{\cc}/\Gamma_{\mathrm{had}}}

\newcommand{\Dspdirect} {\Dsprimep\to\Dstarp\pionp\pionm}
\newcommand{\Dsptodss}  {\Dsprimep\to\Dpzeroone\pionp}
\newcommand{\Dsstods}   {\Dpzeroone\to\Dstarp\pionm}
\newcommand{\Dspviadss} {\mbox{$\Dsptodss$},\ \mbox{$\Dsstods$}}
\newcommand{\threeprong}{\mbox{$\Dstarp\to\Dzero\pionp,\ \Dzero\to\Km\pionp$}}

\newcommand {\downto}
         {\mbox{ \begin{picture}(14,10)
                    \put(0,10){\line(0,-1){5.0}}
                    \put(2,5){\oval(4,4)[bl]}
                    \put(1,0){\makebox(0,0)[bl]{$\rightarrow$}}
                 \end{picture} }}

\begin{document}

\begin{titlepage}

\begin{center}
  {\large   EUROPEAN ORGANIZATION FOR NUCLEAR RESEARCH}
\end{center}
\bigskip

\begin{flushright}
       CERN-EP-2001-003   \\ Journal Version, ..th April, 2001
\end{flushright}
\bigskip\bigskip\bigskip\bigskip\bigskip

\begin{center}
  {\huge\bf A Search for \\
    {a Narrow Radial Excitation \\ \vspace{0.2cm}
     of the \boldmath $\Dstarpm$ \unboldmath Meson}}
\end{center}
\bigskip\bigskip

\begin{center}
{\LARGE The OPAL Collaboration}
\end{center}
\bigskip\bigskip\bigskip

\begin{center}
  {\large\bf  Abstract}
\end{center}

\noindent 
A sample of 3.73 million hadronic $\Z$ decays, recorded with the OPAL
detector at LEP in the years 1991--95, has been used to search for a
narrow resonance corresponding to the decay of the $\Dsprimepm(2629)$
meson into $\Dstarpm\pionp\pionm$. The $\Dstarp$ mesons are
reconstructed in the decay channel $\Dstarp \to \Dzero \pionp$ with
$\Dzero \to \Km\pionp$.  No evidence for a narrow $\Dsprimepm(2629)$
resonance is found. A limit on the production of narrow
$\Dsprimepm(2629)$ in hadronic $\Z$ decays is derived:
\[
    f(\Z \to  \Dsprimepm(2629)) \times Br(\Dsprimep \to \Dstarp \pionp
    \pionm) < 3.1 \times 10^{-3} \ \ \mathrm{(95\%\ C.L.)}
\]

\bigskip\bigskip
\bigskip\bigskip
\bigskip\bigskip

\begin{center}
{\large (Submitted to Euro. Phys. J C)}
\end{center}

\end{titlepage}

\begin{center}{\Large        The OPAL Collaboration
}\end{center}\bigskip
\begin{center}{
G.\thinspace Abbiendi$^{  2}$,
C.\thinspace Ainsley$^{  5}$,
P.F.\thinspace {\AA}kesson$^{  3}$,
G.\thinspace Alexander$^{ 22}$,
J.\thinspace Allison$^{ 16}$,
G.\thinspace Anagnostou$^{  1}$,
K.J.\thinspace Anderson$^{  9}$,
S.\thinspace Arcelli$^{ 17}$,
S.\thinspace Asai$^{ 23}$,
D.\thinspace Axen$^{ 27}$
% ...
}\end{center}\bigskip
\bigskip
$^{  1}$School of Physics and Astronomy, University of Birmingham,
Birmingham B15 2TT, UK
\newline
$^{  2}$Dipartimento di Fisica dell' Universit\`a di Bologna and INFN,
I-40126 Bologna, Italy
\newline
$^{  3}$Physikalisches Institut, Universit\"at Bonn,
D-53115 Bonn, Germany
\newline
% ...
$^{  5}$Cavendish Laboratory, Cambridge CB3 0HE, UK
\newline
% ...
$^{  9}$Enrico Fermi Institute and Department of Physics,
University of Chicago, Chicago IL 60637, USA
\newline
% ...
$^{ 16}$Department of Physics, Schuster Laboratory, The University,
Manchester M13 9PL, UK
\newline
$^{ 17}$Department of Physics, University of Maryland,
College Park, MD 20742, USA
\newline
% ...
$^{ 22}$Department of Physics and Astronomy, Tel Aviv University,
Tel Aviv 69978, Israel
\newline
$^{ 23}$International Centre for Elementary Particle Physics and
Department of Physics, University of Tokyo, Tokyo 113-0033, and
Kobe University, Kobe 657-8501, Japan
\newline
% ...
$^{ 27}$University of British Columbia, Department of Physics,
Vancouver BC V6T 1Z1, Canada
\newline
% ...
Tel Aviv 69978, Israel.

\clearpage\newpage


\section{Introduction}

Quark models based on Quantum Chromodynamics
(QCD)~\cite{dsppred,dspmass} predict the existence of charm mesons
whose radial wave functions are not in their ground state.  A spin
doublet containing a pseudoscalar $\mathrm{2^{1}S_{0}}$ and a vector
$\mathrm{2^{3}S_{1}}$ state is expected. These states are referred to
as the $\Dprime$ and $\Dsprime$, respectively. Their masses are
predicted to be \mbox{$m_{\Dprime}=2.579\GeVcc$} and
\mbox{$m_{\Dsprime}=2.629\GeVcc$} \cite{dspmass} (see
Table~\ref{dspectrum}). The uncertainty on these masses is estimated
to be of the order of $20\MeVcc$ from the difference between the
values obtained in computations \cite{dspmass} and observations
\cite{pdg} of the masses of orbitally excited states.

The kinematically favoured decay mode of the $\Dsprimep$ is presumed
to be the direct three-body decay into a $\Dstarp\pionp\pionm$ final
state\footnote{ Charge conjugates will be implied throughout this
  paper.}. This is an S-wave decay and thus generally expected to be
broad, although a model exists \cite{fluxtube} which estimates the
partial width of this decay mode to be less than $1\MeVcc$.  Decays
via an intermediate orbitally excited $\D$ meson could also contribute
to the same final state; these decay widths are estimated to be
several $\MeVcc$ by the same model. If enough phase space is
available, $\Dsprimep$ decays to a $\Dstarp\rmrho$ state with
$\rmrho\to\pionp\pionm$ might also be allowed.  The decays via
orbitally excited states include S-wave ($\Dspviadss$), and D-wave
transitions (e.g.  \mbox{$\Dsprimep\to\Dstwozero\pionp$},
\mbox{$\Dstwozero\to\Dstarp\pionm$}).  The decays involving a $\rmrho$
are P-wave transitions. The higher partial waves are expected to be
suppressed due to lack of phase space.

\begin{table}[h]
\begin{center}
\begin{tabular}{|c|c||c||c|c|}

\hline

& & predicted mass & \multicolumn{2}{c|}{observed properties \cite{pdg}} \\

\cline{4-5}

\raisebox{1.5ex}[-1.5ex]{state}          &
\raisebox{1.5ex}[-1.5ex]{$J_j^P$}        &
$\left[ {\GeVcc} \right]\!$  \cite{dspmass}&
\small mass $\left[ {\GeVcc} \right]\!$  &
\small width $\left[ {\MeVcc} \right]\!$ \\

\hline
$\D$        & $0_{1/2}^{-}$ & 1.875 & 1.865 &  \\
$\Dstar$    & $1_{1/2}^{-}$ & 2.009 & 2.010 & $<0.131$ \\
$\Dstar_0$  & $0_{1/2}^{+}$ & 2.438 &       &  \\
$\Dstar_1$ & $1_{1/2}^{+}$ & 2.501 & \raisebox{1.0ex}[-1.0ex]{\Huge \}} \raisebox{1.5ex}[-1.5ex]{ (2.461) \cite{cleodss}}      & \raisebox{1.5ex}[-1.5ex]{(290) \cite{cleodss}} \\
$\D_1$      & $1_{3/2}^{+}$ & 2.414 & 2.422  & 18.9 \\
$\Dstar_2$  & $2_{3/2}^{+}$ & 2.459 & 2.459  & 23   \\
$\Dprime$   & $0_{1/2}^{-}$ & 2.579 &        &      \\
$\Dsprime$  & $1_{1/2}^{-}$ & 2.629 & (2.637) \cite{delphidsp} & ($<15$) \cite{delphidsp} \\
\hline
\end{tabular}
\caption{\label{dspectrum} Masses of charm mesons as predicted in
  Ref.~\cite{dspmass}. Measurements of the meson masses and width are
  given where available~\cite{pdg}. Values in brackets are not from
  Ref.~\cite{pdg}, but represent recent observations from
  CLEO~\cite{cleodss} and DELPHI~\cite{delphidsp}. $\Dstar_0$,
  $\Dstar_1$, $\D_1$, and $\Dstar_2$ are orbital excitations (the
  $\Dstar_1$ meson is also known as $\Dprime_1$). $\Dprime$ and
  $\Dsprime$ are radial excitations.}
\end{center}
\end{table}

The only experimental evidence for the existence of radially excited
$\D$ mesons comes from the DELPHI collaboration which has published an
observation of a narrow resonance decaying to $\Dstarp\pionp\pionm$
\cite{delphidsp}. The observed mass of
$2.637\pm0.002\mathrm{(stat.)}\pm0.006\mathrm{(syst.)}\GeVcc$ is close
to the theoretical prediction. However, the interpretation of this
$\D$ resonance as $\Dsprime$ is controversial, since the narrow width
is incompatible with most theoretical models \cite{theoryresp}.  The
DELPHI measurement of the width of the state is limited by their
detector resolution and quoted to be smaller than $15\MeVcc$ at $95\%$
C.L.

This paper presents a search for the state reported by DELPHI in
hadronic $\Z$ decays recorded with the OPAL detector.  The search
covers both the region of the resonance measured by DELPHI as well as
any other narrow resonance in the vicinity of the predicted
$\Dsprimepm$ mass.  Therefore the particle being searched for will be
referred to as $\Dsprimepm(2629)$ throughout this paper.

After a description of the data and Monte Carlo samples, the selection
criteria for the $\Dsprimep$ mesons are presented. Since the
production mechanism of $\Dsprimep$ mesons is different in $\Z \to
\bb$ and $\Z \to \cc$ events, two different selections are used to
reconstruct $\Dsprimep$ mesons. One is optimized to select $\Dsprimep$
candidates in primary b-events using vertex information, while the
second selection uses harder cuts on the momentum of the $\Dsprimep$
candidate, resulting in a sample enhanced in $\cc$ events. In Section
\ref{results}, the selection results for data and Monte Carlo events
are presented. Systematic checks are discussed subsequently, and the
calculation of limits on the production rate is described in Section
\ref{limits}. In the concluding Section our results are compared with
those published by DELPHI \cite{delphidsp}.


\section{The OPAL Detector and Event Simulation}
\label{simulation}
       
A detailed description of the OPAL detector can be found elsewhere
\cite{bib-OPALdetector}. The most important components of the detector
for this analysis are the silicon microvertex detector, the tracking
chambers and the electromagnetic calorimeter. The microvertex detector
consists of two layers of silicon strip detectors which provide high
spatial resolution near the interaction region. The central jet
chamber is optimized for good spatial resolution in the plane
perpendicular to the beam axis\footnote{The OPAL coordinate system is
  defined as a right-handed Cartesian coordinate system, with the
  $x$-axis pointing in the plane of the LEP collider towards the
  centre of the ring and the $z$-axis along the electron beam
  direction.}.  The resolution along the beam direction is improved by
the $z$~information delivered by the silicon microvertex detector, by
a vertex drift chamber between the silicon detector and the main tracking
chambers, and by $z$-chambers surrounding the main tracking chamber.
The central detector provides precise determination of momenta of
charged particles by measuring the curvature of their trajectories in
a magnetic field of $0.435\;$T. The solenoid is mounted outside the
tracking chambers but inside the electromagnetic calorimeter, which
consists of approximately $12\thinspace 000$~lead glass blocks
providing an azimuthal coverage up to polar angles of $|\cos
\theta|<0.98$. To improve the shower energy and spatial resolution, a
preshower detector (presampler) is mounted between the solenoid and
the electromagnetic calorimeter.

For background studies, about 8 million hadronic decays of the $\Z$
have been simulated using the JETSET 7.4 Monte Carlo \cite{bib-JETSET}
with parameters tuned to the OPAL data \cite{bib-OPALtune}. In
addition, samples with simulated $\Dsprime$ decays have been generated
since JETSET does not include $\Dsprime$ production by default.  For
the production of the $\Dsprime$ Monte Carlo samples with JETSET,
radially excited $\D$ meson states have been implemented into JETSET
with the mass set to $2.629\GeVcc$, according to the most recent
calculation \cite{dspmass}, and the width set to zero. The possible
effects of a non-zero width are treated later as systematic
uncertainty.  DELPHI only measured the ratio of the $\Dsprimep$ and
$\Dsszero$ production rates\footnote{The symbol $\Dss$ represents the
  two narrow orbital resonances $\Done$ and $\Dstwo$.}
\cite{delphidsp}, and therefore this measurement has been combined
with the OPAL measurement of $\Dsszero$ production \cite{opaldss} to
calculate the expected absolute $\Dsprimep$ production rate.
$\Dsprime$ production is simulated in fragmentation and in $\B$ meson
decays. In the latter case, $\Dsprime$ mesons were produced in the
same b decay channels as $\D_2^{*}$, in proportion chosen to reproduce
the DELPHI measurement \cite{delphidsp}.

Four samples of $\Dsprime$ Monte Carlo events have been generated:
production in $\Z\to\cc$ and $\Z\to\bb$ events and the decay channels
$\Dspdirect$ and $\Dsptodss$. The signal Monte Carlo sample contains
600,000 events in total, with at least one charged $\Dsprime$ meson in
each event.  In all Monte Carlo samples, heavy quark fragmentation has
been implemented using the the model of Peterson
et~al.~\cite{bib-PETERSON} with fragmentation parameters determined
from LEP data \cite{bib-LEPNIM}.  The samples have been passed through
a detailed simulation of the OPAL detector \cite{bib-OPALGOPAL} and
then analyzed in the same manner as the data.


\section{Event and Track Selection}
\label{tracksel}

Hadronic $\Z$ decays are selected based on the number of reconstructed
tracks and the energy deposited in the electromagnetic calorimeter, as
described in Ref.~\cite{bib-OPALmh}.  The analysis uses an initial
sample of $3.73$~million hadronic decays of the $\Z$ collected with
the OPAL detector between 1991 and 1995.

Tracks are used in the reconstruction if they pass loose quality cuts
requiring $|d_0| < 0.5\cm$, $|z_0|<20\cm$, $p>0.5\GeVc$, and $p_{xy}>
0.150\GeVc$.  Here $|d_0|$ is the distance of closest approach of the
track to the primary vertex, measured in the plane perpendicular to
the beam, $z_0$ the distance along the beam at this point, $p$ the
momentum, and $p_{xy}$ the momentum in the plane perpendicular to the
beam. Only tracks with more than 40 hits out of a maximum of 159 in
the main tracking chamber are used.  The primary vertex in a collision
is reconstructed from the tracks in the event and constrained by the
known average position and spread of the $\ee$ interaction point. To
improve the mass resolution, a track is required to have a precise $z$
measurement at the exit point of the tracking chamber, either from an
associated hit in the $z$-chamber surrounding the central drift
chamber, or from the presampler. In cases where the particle exited
the tracking chamber in the endcap region, i.e.~$|\cos \theta|>0.73$,
the exit point is determined precisely from the last sense wire used
for the track measurement.


\section{Reconstruction of \boldmath $\Dsprimep$ \unboldmath Candidates}

The reconstruction of $\Dspdirect$ decays involves $\Dstarp$
reconstruction and, subsequently, the combination of accepted
$\Dstarp$ candidates with additional pion candidates.  The $\Dstarp$
reconstruction follows closely a recent OPAL publication
\cite{rcpaper}, with the exception of the cut on the kaon particle
identification probability which has been tightened for this analysis,
in order to reduce combinatorial background.

The following decay sequence is reconstructed exclusively:
\begin{center}
\parbox{6cm}{
\begin{tabbing}
  $\Dsprimep\to$\= \downto \= xxx \kill
  $\Dsprimep\to$\=$\Dstarp\pionp\pionm$\\
                \> \downto $\Dzero\pionp$ \\
                \>         \> $\downto\Km\pionp$ \\
\end{tabbing}
}
\end{center}
All tracks that fulfil the criteria described in
Section~\ref{tracksel} are considered for the construction of
$\Dsprimep$ candidates. Only combinations of tracks that have the
correct charge assignments, and which pass intermediate requirements
imposed by the reconstruction of the $\Dstarp$, are taken into
account. The two pions produced in the initial $\Dsprimep$ decay will
be referred to as ``$\Dsprimep$ pions'', whereas the pion coming from
the $\Dstarp$ decay will be referred to as ``slow pion''.

Two tracks of opposite charge are accepted as a $\Dzero$ candidate if
their invariant mass lies within the range $1.79 - 1.94\GeVcc$,
assigning the pion mass to one particle and the kaon mass to the other
\cite{pdg}.  $\Dzero$ candidates are combined with tracks of charge
opposite to that of the kaon candidate to form $\Dstarp$ candidates.
The scaled energy $\xE$ of the $\Dstarp$ candidate, i.e.~the ratio of
the energy sum of all participating tracks over the beam energy, is
required to be larger than $0.2$ in order to reject combinatorial
background from low-energy fragmentation tracks. The difference of the
invariant masses of $\Dstarp$ and $\Dzero$ candidates,
$\dm=\mDstar-\mDzero$, must lie within the interval from $142$ to
$149\MeVcc$.

Background in the sample is further reduced by cutting on the helicity
angle $\theta^*$ measured between the direction of the $\Dzero$
candidate in the laboratory frame and the direction of the kaon in the
rest frame of the $\Dzero$ candidate. The kaon candidate from the
$\Dzero$ decay is expected to be isotropically distributed in $\cos
\theta^*$, while the background peaks at $\cos \theta^* = -1$ and,
particularly at small $\xE$, at $\cos \theta^* = +1$.  This effect is
taken into account by requiring $\costs>-0.9$ for $\xE>\mathrm{0.5}$
and $|\costs|<0.8$ for $\xE<\mathrm{0.5}$.

At low $\xE$, where the background is most pronounced, the particle
identification power of the OPAL detector is used to enrich the sample
in true kaons from $\Dzero$ decays. A probability $W_{\K}$ is computed
from the ionization energy loss measurement of a track in the drift
chamber, the track momentum, and the theoretical expectation for a
kaon. At least $20$ $\dEdx$ out of a possible 159 measurement points
and $W_{\K}>10\%$ are required for the kaon candidate in a $\Dstarp$
candidate with $\xE<\mathrm{0.5}$.

A fraction ($1.5\%$) of all $\Dstarp$ candidates share the same slow
pion candidate with another $\Dstarp$ candidate in the same event. In
this case only the $\Dstarp$ candidate with a $\Dzero$ invariant mass
closest to the mean value obtained from a fit to the $\Dzero$ mass
distribution is kept for the further analysis.  The $\Dstarp$
candidates surviving the cuts described above are combined with all
remaining pairs of oppositely charged tracks.  For these combinations,
a mass difference $\dmp=\mDsp-\mDstar$ is calculated.

\begin{figure} [tb]
  \center{ \epsfig{file=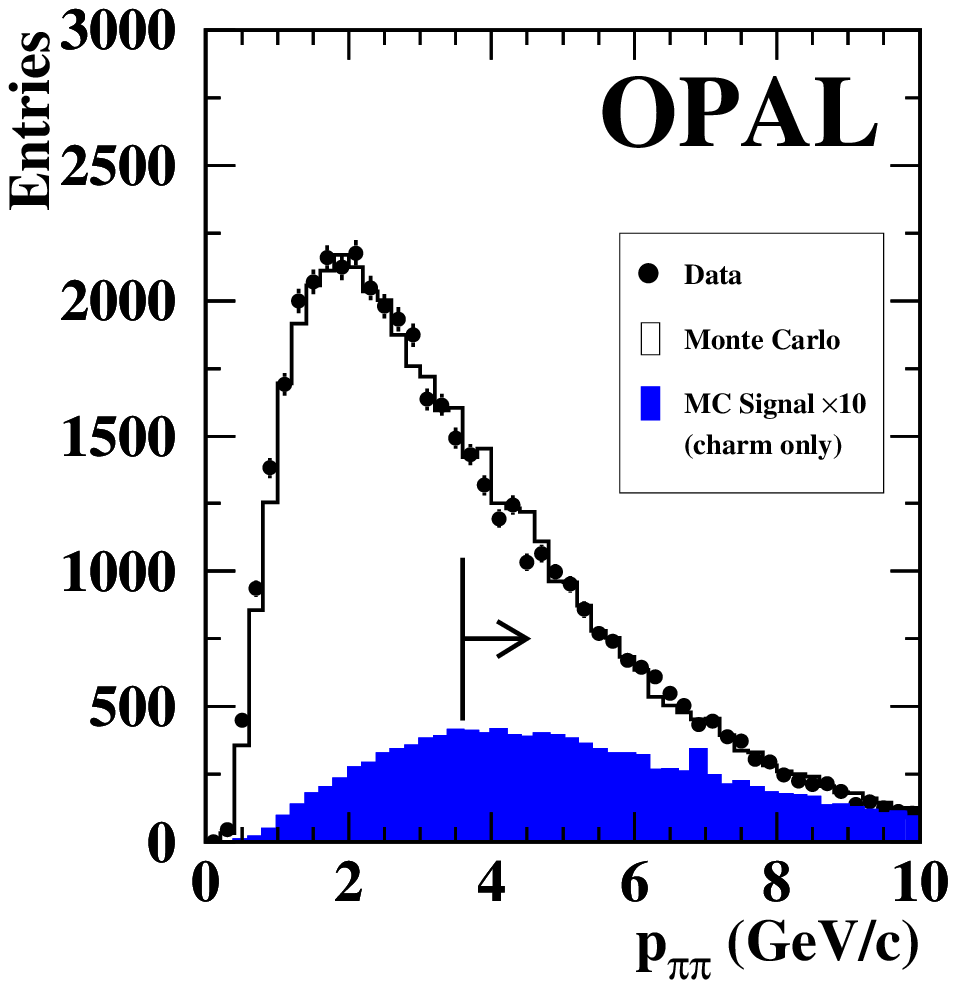, height=8cm}
    \smcap{\label{pimom2} Magnitude of the vector sum of the
      $\Dsprime$ pion momenta, p$_{\pi\pi}$, for data (points with
      error bars), Monte Carlo without $\Dsprimep$ (open histogram,
      scaled to the same number of entries as data) and true
      $\Dsprimep$ (shaded, scaled up by a factor of 10). The arrow
      indicates the selected region.}  }
\end{figure}

In order to suppress background from $\uds$ events, charm and bottom
tags are used. The mean fractional energy of $\Dsprimep$ mesons in
$\Z\to\cc$ events is expected to be large compared to $\Dsprimep$ from
$\B$ decays, and especially compared to fake $\Dsprimep$ candidates in
events of all flavours consisting of low-energy fragmentation tracks.
Since most of the energy of the $\Dsprimep$ is transferred to the
$\Dstarp$, a charm enriched sample is selected by imposing a cut on
the energy fraction of the $\Dstarp$ of $\xE>\mathrm{0.4}$.
Additionally, in the Monte Carlo simulation, the $\Dsprimep$ pions
have a higher momentum if the $\Dsprime$ is produced in $\cc$ events
compared to the production via the decay of b hadrons.  This
information has been exploited by selecting only candidates for which
the magnitude of the vector sum of the two $\Dsprime$ pion momenta is
greater than $3.6\GeVc$ (Figure~\ref{pimom2}). The mass resolution in
the charm enriched sample is improved by constraining all $\Dsprimep$
candidate tracks except those from the $\Dzero$ decay to the primary
event vertex.

A bottom-enriched sample is selected by requiring an apparent $\Dzero$
decay length of at least $0.3\mm$, defined as the distance between
the reconstructed $\Dzero$ vertex and the primary event vertex,
measured in the plane perpendicular to the beam.  $\Dzero$ candidates
from light quark ($\uds$) events and $\cc$ events are expected to be
made up from tracks in the vicinity of the primary vertex. In
contrast, $\Dzero$ candidates in $\bb$ events tend to have a larger
decay length with respect to the primary vertex, due to the $\B$ meson
lifetime.

\begin{figure} [tb]
  \center{ \epsfig{file=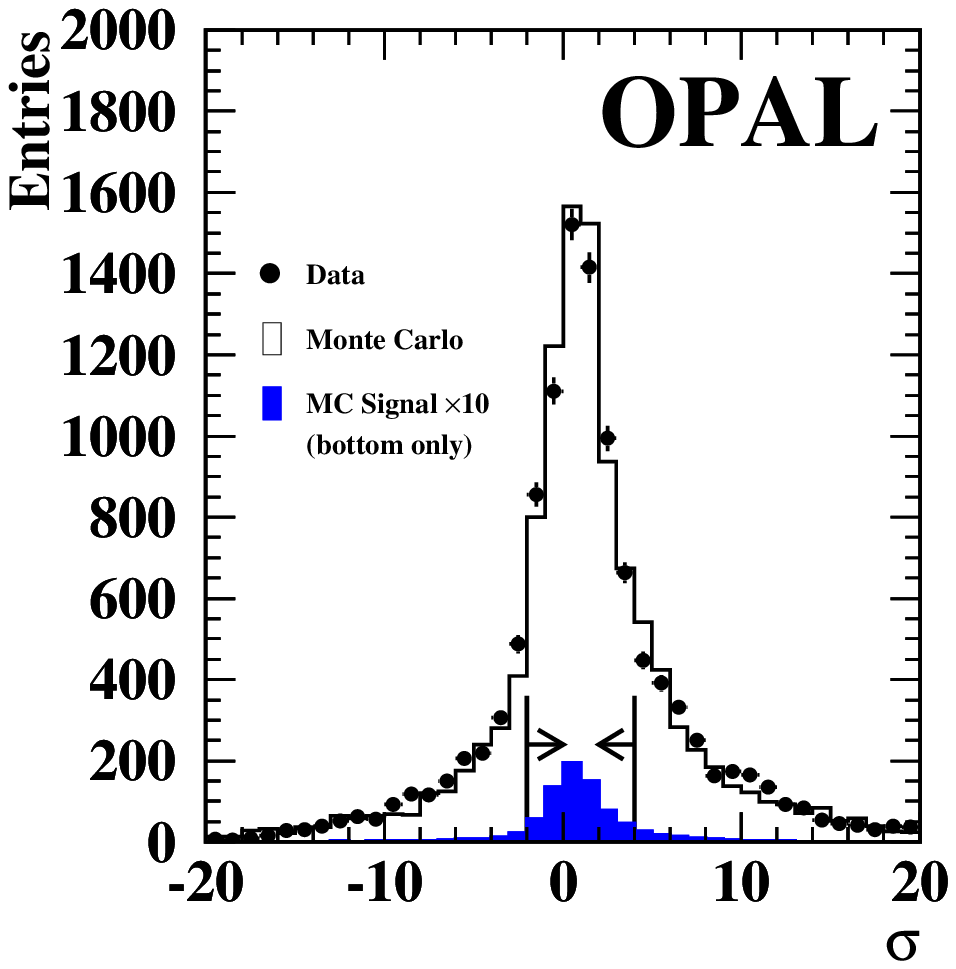, height=8cm} \smcap{\label{vsign2}
      Separation significance $\stddev$ between the $\Dzero$ vertex
      and the reconstructed vertex of the $\Dsprimep$ decay pion
      tracks for data (points with error bars), Monte Carlo (open
      histogram, scaled to the same number of entries as data) and
      true $\Dsprimep$ (shaded, scaled up by a factor of 10) for
      candidates passing all selection criteria except for the cut on
      the quantity shown here. The arrows indicate the selected
      region.}  }
\end{figure}

The separation significance of two vertices is defined as the distance
between them in the plane perpendicular to the beam, divided by the
uncertainty on this quantity. The separation significance of the
$\Dzero$ vertex and the reconstructed vertex formed by the $\Dsprime$
pion tracks is required to be between $-2.0$ and $+4.0$. The
distribution of this variable is not centred at zero, because the
$\Dzero$ has a lifetime large enough to be observed.  The separation
significance between the $\Dsprimep$ pion vertex and the $\Dzero$
vertex is shown in Figure~\ref{vsign2}.  A positive sign indicates
that the $\Dzero$ vertex is farther from the primary interaction point
than the $\Dsprimep$ pion vertex. Positive values are expected for
signal, because the $\Dzero$ occurs later in the decay chain.

\begin{table}[h]
  \begin{center}
  \begin{tabular}{|l|c|c|c|}
  \hline
  flavour  & charm enriched & bottom enriched & combined sample \\
  \hline
  \hline
  $\uds$   & $12\%$       & $10\%$        & $11\%$          \\
  $\qc$    & $60\%$       & $15\%$        & $31\%$          \\
  $\qb$    & $28\%$       & $75\%$        & $58\%$          \\
  \hline
  \end{tabular}
  \smcap{\label{taggedcand} Flavour composition of the charm enriched,
  bottom enriched and combined Monte Carlo samples.}
  \end{center}
\end{table}

The Monte Carlo flavour composition of the $\Dsprimep$ candidates
passing all selection criteria is shown in Table~\ref{taggedcand}.
For $\Dsprimep$ mesons from primary charm and bottom quarks, the
reconstruction efficiencies are estimated from Monte Carlo analysis to
be $14.1\%$ and $7.1\%$, respectively.  The combined efficiency for
$\Dsprimep$ from both sources is $11.4\%$, assuming equal rates for
production in charm and bottom events.  This assumption is motivated
by the range of production rate ratios observed for $\Dstarpm$
\cite{rcpaper} and orbitally excited $\Dsszero$ mesons \cite{opaldss}.
The analysis has also been performed with several other assumptions,
and the dependence of the results on these different scenarios is
included as a systematic error.


\section{Selection Results}
\label{results}

In order to maximize the expected sensitivity of the analysis, the
charm and bottom enriched samples with 1765 and 3051 candidates,
respectively, in the $\dmp={\mathrm m}_{\Dsprime}-{\mathrm
  m}_{\Dstar}$ region between $2.3\GeVcc$ and $3.0\GeVcc$ are merged
into a combined sample. The overlap of 324 candidates between charm
and bottom sample is taken into account; candidates that have been
tagged as charm and bottom are used only once. This sample contains a
total of $4492$~$\Dsprimep$~candidates in 2192 events. The results for
the combined sample and the $\qb$ and $\qc$ enriched samples are
discussed below.

\begin{figure} [htb]
  \center{ \epsfig{file=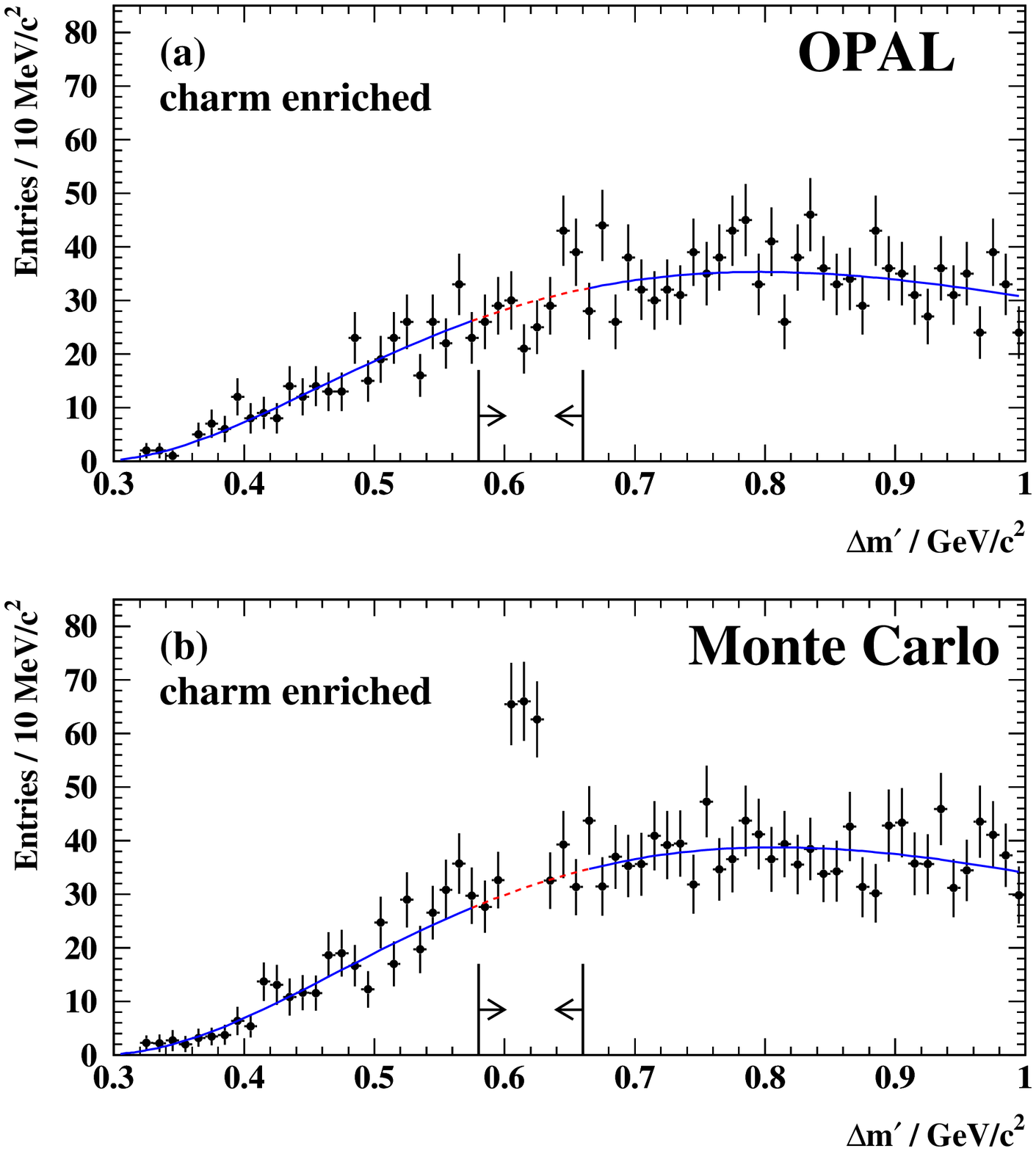,height=19cm}
    \smcap{\label{mpdistrc} Distribution of the mass difference
      $\dmp=\mDsp-\mDstar$ of $\Dsprimep$ candidates in the charm
      enriched sample, for (a) data and (b) Monte Carlo. The
      $\chi^2$/dof of the background fit is 0.80 for data, and 0.70
      for the Monte Carlo simulation.}}
\end{figure}

\begin{figure} [htb]
  \center{ \epsfig{file=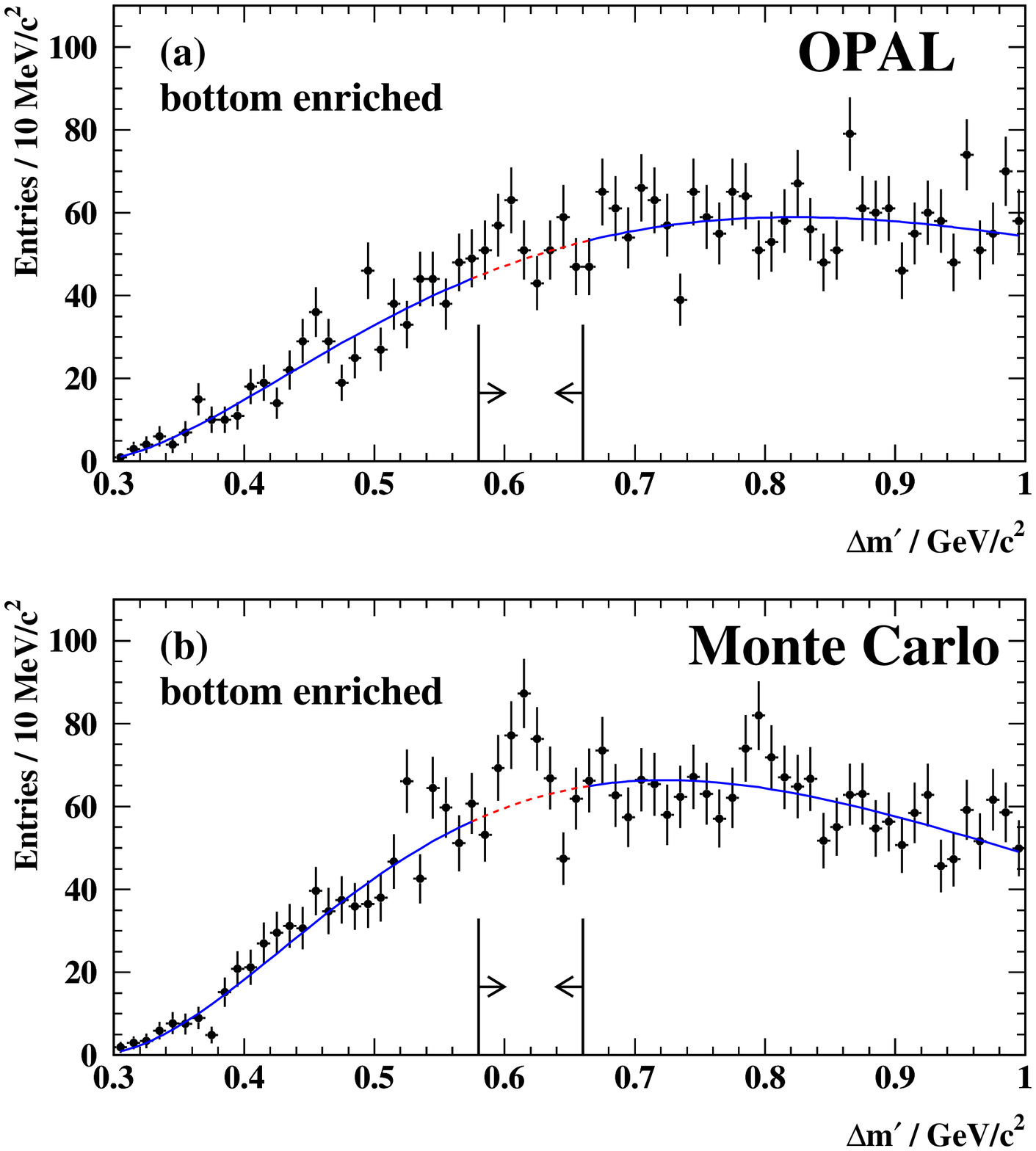,height=19cm}
    \smcap{\label{mpdistrb} Distribution of the mass difference
      $\dmp=\mDsp-\mDstar$ of $\Dsprimep$ candidates in the bottom
      enriched sample, for (a) data and (b) Monte Carlo. The
      $\chi^2$/dof of the background fit is 1.20 for data, and 1.07
      for the Monte Carlo simulation. The difference in shape and
      normalization of the background is explained by a large number
      of partially reconstructed $\Dsprimep$ candidates which, due to
      their low mean scaled energy, mainly contribute to the b
      enriched sample. The analysis has also been performed on a Monte
      Carlo sample without $\Dsprime(2629)$ production.  It has been
      found that in this case the background shape and normalization
      agree very well with the data.}}
\end{figure}

The $\dmp$ distribution of the combined charm and bottom enriched
sample is shown in Figure~\ref{mpdistr}a. No narrow resonance is
observed anywhere in the $\dmp$ region between $0.3$ and $1.0\GeVcc$,
although the Monte Carlo simulation with a $\Dsprime$ production rate
fixed at the value published by DELPHI~\cite{delphidsp} shows a clear
signal (see Figure~\ref{mpdistr}b). This is also separately true for
the $\qc$ and $\qb$ enriched samples, shown in Figures~\ref{mpdistrc}
and~\ref{mpdistrb}, respectively. Shape and normalization of the
background differ in data and simulation, especially in the b enriched
sample. This is due to $\Dsprimep$ candidates in the Monte Carlo where
one track is incorrectly assigned to the $\Dsprimep$ candidate.  These
candidates lead to a broad enhancement in the mass distribution.  The
mean scaled energy $\xE$ of these candidates is small. Therefore, they
contribute more to the b enriched sample than to the c enriched
sample.  The analysis has also been performed on a Monte Carlo sample
without $\Dsprime(2629)$ production. It has been found that in this
case the background shape and normalization agree very well with the
data.  This absence of a significant enhancement over the expected
background in data provides additional confidence that the
$\Dsprimep(2629)$ production rate must be small.

A limit on the $\Dsprimep(2629)$ production rate is calculated by
defining a mass window in the $\dmp$ range of
$0.58\GeVcc<\dmp<\mathrm{0.66}\GeVcc$, corresponding to a $\Dsprime$
mass window of \mbox{$2.59-2.67\GeVcc$}.  This includes the
$\pm2\stddev$ range of both the theoretical prediction of the
$\Dsprime$ \cite{dspmass} and the excess observed by DELPHI
\cite{delphidsp}.  The background distribution has been fitted using a
parametrisation of the form \cite{delphidsp,opaldss}
\begin{equation}
\mathrm{f}(\dmp)=\mathit{\alpha}\;
(\mathit{\dmp -m_{\mathrm{0}}})^{\mathit{\beta}}\;
\mathrm{e}^{-\mathit{\gamma}(\mathit{\dmp -m_{\mathrm{0}}})}\ ,
\end{equation}
where $\mathit{m}_{\mathrm{0}}=\mathrm{0.28}\GeVcc$ is the kinematic
limit for $\Dstarp\pionp\pionm$ combinations, and $\alpha$, $\beta$
and $\gamma$ are the fit parameters. The signal region has been
excluded from the fit. The number of $\Dsprimep$ candidates in the
mass window is determined to be $14\pm28$ by subtracting the
integrated interpolated background function from the observed number
of candidates in the signal interval.  The quoted statistical error is
obtained by combining the Gaussian variance of the number of entries
in the mass window ($\pm25$) and the uncertainty on the background
integral obtained by propagating the uncertainties on the fit
parameters ($\pm14$), taking into account the correlations.


\section{Systematic Checks}
\label{systematics}

\begin{figure} [htb]
  \center{ \epsfig{file=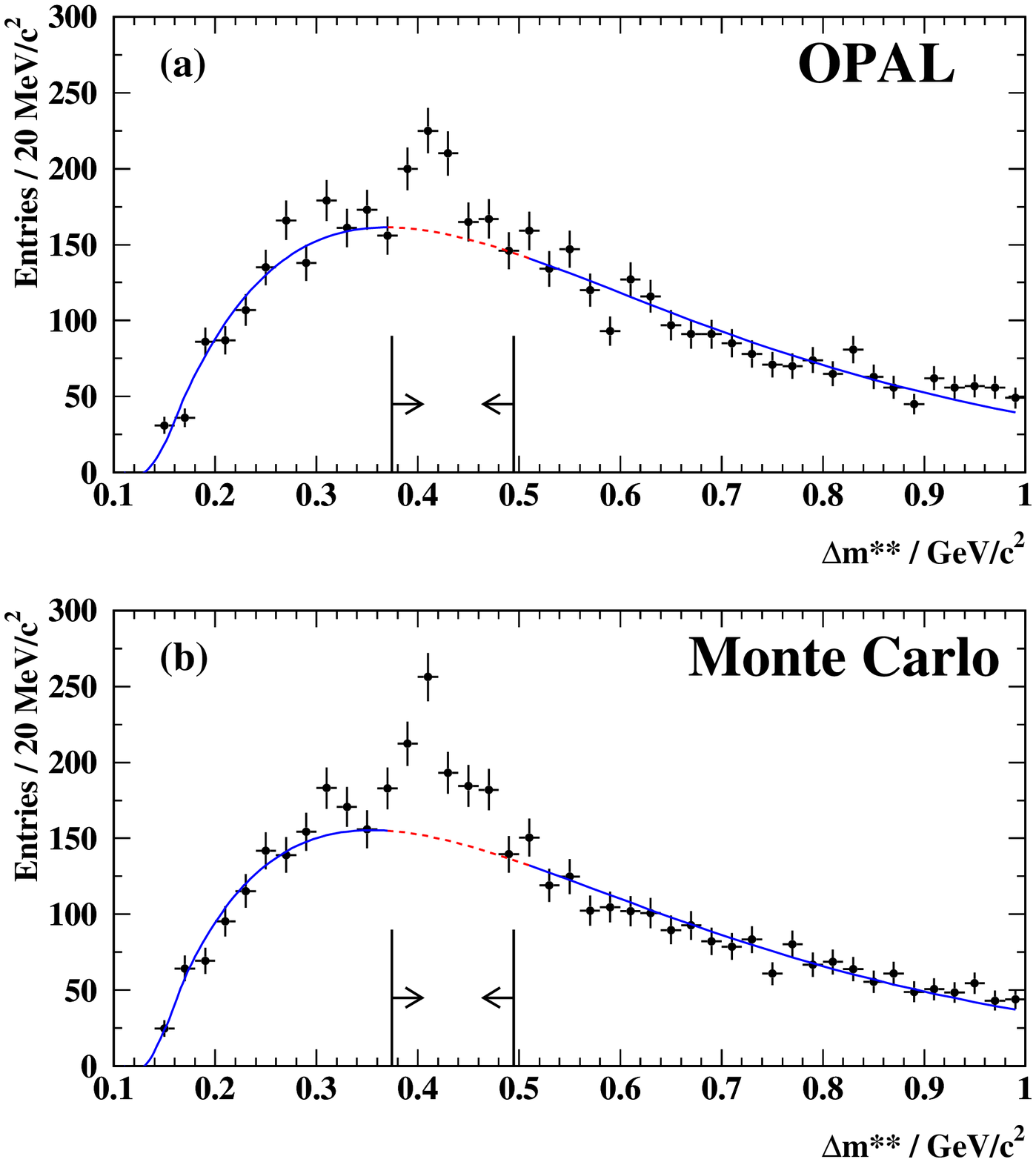,height=19cm}
    \smcap{\label{dssmass} Distribution of the mass $\Delta m_{**} =
      m_{\Dstarp\pionm}-m_{\Dstarp}$, for (a) data and (b) Monte
      Carlo.  The production rates of $\Dsszero$ mesons in the Monte
      Carlo have been adjusted to the values measured in data
      \cite{opaldss}.  The same parametrisation as in the $\Dsprimep$
      analysis is used to describe the background.  The $\chi^2$/dof
      of the fit is 1.43 for data, and 1.22 for the Monte Carlo
      simulation.  The Monte Carlo histogram is scaled to the number
      of hadronic events in data.}  }
\end{figure}

The $\Dsprimep$ analysis has been checked by applying a similar
selection to $\Dstarp\pionm$ final states, looking for the narrow
orbital resonances $\Dzeroone$ and $\Dstwozero$. The $\Dstarp$
selection criteria are as described earlier. Accepted $\Dstarp$
candidates are combined with pion candidates passing selection
criteria similar to those applied to the $\Dsprimep$ pions, except for
the cuts on the pion-pion momentum sum and the pion-pion vertex
separation, which are inappropriate here.  Instead, the pion momentum
is required to be larger than 2.0 GeV/c, as was done in an earlier
OPAL publication on orbitally excited D mesons \cite{opaldss}.  The
efficiency of the $\Dsszero\to\Dstarp\pionm$ reconstruction is about
$7.3\%$. The results achieved on data and simulated events can be
compared in Figure \ref{dssmass}.  The method used for determining the
number of signal events is the same as was used for the $\Dsprimep$.
The lower mass window boundary has been set to $2.382\GeVcc$, two
$\Dzeroone$ widths below the world average $\Dzeroone$ mass. The upper
boundary has been chosen to be $2.507\GeVcc$, two $\Dstwozero$ widths
above the world average $\Dstwozero$ mass.  The production rates of
the two narrow $\Dsszero$ resonances have been adjusted in the
simulation to match the previous OPAL measurement \cite{opaldss}.  The
amount of signal found in Monte Carlo is $281\pm16$, whereas in data
the excess is $189\pm41$ in a sample of 4711 selected $\Dsszero$
candidates.  The analysis presented here and the previous OPAL
publication on orbitally excited D mesons \cite{opaldss} that was used
to adjust the $\Dsszero$ production rates in Monte Carlo use a similar
dataset. Both results are therefore expected to be correlated.
Nevertheless, it has been found that the overlap of the selected
$\Dsszero$ candidate samples of old and new analysis is very small,
and the results can thus be considered almost statistically
independent. This leads to the conclusion that the measured rates are
statistically compatible.

Having confirmed in the case of $\Dstarp\pionm$ final states that the
reconstruction procedure performs as expected, the results of the
search in the $\Dstarp\pionp\pionm$ final state are used to place
limits on the production rate of the $\Dsprimep(2629)$.  The
efficiency of $\Dsprimep$ reconstruction is taken from Monte Carlo
simulation.  The efficiency calculation accounts for several sources
of systematic uncertainties. They are evaluated as follows:
\begin{itemize}
\item The efficiency has been calculated from a simulation with zero
  $\Dsprimep$ width. To quantify the effect of a non-zero width, the
  analysis has been repeated with a $\Dsprimep$ width of $15\MeVcc$,
  the $95\%$ C.L. upper limit obtained by DELPHI \cite{delphidsp}.
  This results in a relative decrease of the efficiency of $16.5\%$,
  taken as systematic uncertainty.
\item The simulated detector resolutions of momentum, impact
  parameters and track angles have been varied by $\pm10\%$, as in
  previous publications \cite{opaldss,rcpaper}, around the values that
  best describe the data. The largest relative deviation in the
  selection efficiency ($^{+4.9\%}_{-5.7\%}$) is taken as systematic
  uncertainty.
\item Equal production rates of $\Dsprimep$ in $\qc$ and $\qb$ decays
  have been assumed. Other possible assumptions include the same ratio
  as observed in $\Dstarpm$ production \cite{rcpaper}, or in
  $\Dsszero$ production \cite{opaldss}. Equal numbers of $\Dsprime$
  from both primary heavy flavours are simulated as well as the ratio
  of candidates from $\bb$ and $\cc$ events observed by DELPHI
  \cite{delphidsp}, corrected for the different reconstruction
  efficiencies of the respective analyses. The ratios thus obtained
  cover a $^{+33\%}_{-23\%}$ range around the central assumption of
  equal production rates in $\bb$ and $\cc$ events. The largest
  observed variations in the efficiency in both directions,
  $^{+3.8\%}_{-4.6\%}$, are taken as systematic uncertainty.
\item A Peterson function has been used to parametrise the heavy quark
  fragmentation in the Monte Carlo samples.  Other fragmentation
  models by Collins and Spiller \cite{fragcs} and Kartvelishvili
  \cite{fragk} have also been tested.  The fragmentation parameters
  have been varied in the limits implied by the measurement of the
  mean $\xE$ distribution of $\Dstarp$ mesons at LEP
  \cite{bib-LEPNIM}. The largest resulting uncertainty of
  $^{+1.7\%}_{-2.3\%}$ is taken as systematic uncertainty.
\item The efficiency depends on the performance of the particle
  identification by $\dEdx$ measurements.  The $\dEdx$ calibration for
  kaons has been checked in a previous analysis \cite{gluoncc} under
  identical circumstances. An error of $\pm3.2\%$ was found for the
  total rate of kaons passing the selection. Because this cut is
  applied only for candidates with $\xE<\mathrm{0.5}$, the expected
  contribution to the uncertainty on the efficiency is $\pm2.1\%$.
\item The $\Dzero$ lifetime is currently measured with a precision of
  $0.7\%$, whereas the average b hadron lifetime is known to
  $0.9\%$~\cite{pdg}. The cut on the $\Dzero$ apparent decay length
  and thus the $\qb$ flavour enrichment is sensitive to the modelling
  of this quantity. The cut value has therefore been varied within the
  combined uncertainty on $\Dzero$ and $\B$ lifetime. The resulting
  deviation of $^{+0.2\%}_{-0.0\%}$ is taken as systematic
  uncertainty.
\item The calculated production limit depends on the branching ratio
  $\mathrm{Br}(\Dstarp\to\Dzero\pionp) \times
  \mathrm{Br}(\Dzero\to\Km\pionp)$. The world average value is
  $(0.677\pm0.005)\times(0.0383\pm0.0009)$~\cite{pdg}. The relative
  uncertainty on the product branching ratio is $\pm2.5\%$.
\item The reconstruction efficiencies of the decay chains
  \mbox{$\Dsptodss$}, \mbox{$\Dsstods$} and $\Dspdirect$ are identical
  within the statistical uncertainties. No systematic uncertainty is
  introduced.
\item The reconstructed number of $\Dsprimep$ in Monte Carlo has been
  compared to the actual number of $\Dsprimep$ in the sample. The
  excess over the background fit is $138\pm30$, with $175$ entries due
  to $\Dsprimep$ with all tracks reconstructed.  It is thus concluded
  that the fit is free of significant bias.
\item The width of the signal region has been varied from $90\%$ to
  $130\%$ of the nominal value to check whether the resulting counting
  rate variations are consistent with statistical effects.  All
  results are clearly within the statistical error.  Thus no
  additional systematic uncertainty is assigned.
\item The bin width of the $\dmp$ histograms in which the fit is
  performed has been varied from $75\%$ to $125\%$ of the value used
  for the analysis. No significant impact on the rate is observed.
\end{itemize}
The contributions to the systematic uncertainty are summarized in
Table~\ref{systerr}.

\begin{table}[tb]
  \renewcommand{\arraystretch}{1.3}
  \center{\begin{tabular}{|l|r|r|r|}
  \hline
  error source & \multicolumn{3}{|c|}{relative contribution} \\
               & $\qc$ enriched & $\qb$ enriched & combined \\
  \hline
  \hline
  \multicolumn{4}{|l|}{\it relative errors on Monte Carlo efficiency} \\
  \hline
  $\Dsprimep$ width
  &  $^{\ +0.0\%}_{-17.9\%}$ & $^{\ +0.0\%}_{-10.6\%}$ & $^{\ +0.0\%}_{-16.5\%}$ \\

  detector resolution              
  &  $^{+5.2\%}_{-6.1\%}$ &  $^{+4.5\%}_{-5.7\%}$ &  $^{+4.9\%}_{-5.7\%}$ \\

  relative production rates in $\qb$ and $\qc$
  &  none                 &  none                 &  $^{+3.8\%}_{-4.6\%}$ \\

  fragmentation modelling
  &  $^{+3.6\%}_{-3.7\%}$ &  $^{+1.1\%}_{-0.3\%}$ &  $^{+1.7\%}_{-2.3\%}$ \\

  kaon $\dEdx$                     
  &  $^{+1.3\%}_{-1.3\%}$ &  $^{+2.6\%}_{-2.6\%}$ &  $^{+2.1\%}_{-2.1\%}$ \\

  $\B$ and $\Dzero$ lifetimes                     
  &  none                 &  $^{+0.5\%}_{-0.0\%}$ &  $^{+0.2\%}_{-0.0\%}$ \\

  Monte Carlo statistics           
  &  $^{+0.1\%}_{-0.1\%}$ &  $^{+0.08\%}_{-0.08\%}$ &  $^{+0.07\%}_{-0.07\%}$ \\

  \hline
  total
  & $^{\ +6.5\%}_{-19.3\%}$ & $^{\ +5.3\%}_{-12.3\%}$ & $^{\ +6.8\%}_{-18.4\%}$ \\
  \hline
  \hline
  \multicolumn{4}{|l|}{\it relative errors on external branching ratios} \\
  \hline
  branching ratio $\threeprong$
  &  $\pm2.5\%$           &  $\pm2.5\%$           &  $\pm2.5\%$ \\
  error on $\GGbb$ and $\GGcc$
  &  $\pm2.9\%$           &  $\pm0.3\%$           &  none    \\
  \hline
  \hline
  total
  & $^{\ +7.5\%}_{-19.7\%}$ & $^{\ +5.9\%}_{-12.6\%}$ & $^{\ +7.2\%}_{-18.6\%}$ \\
  \hline
  \end{tabular}
  \smcap{\label{systerr} Overview of the systematic error sources
  contributing to the total uncertainty. }
  }
\end{table}


\section{Calculation of \boldmath $\Dsprimepm(2629)$ \unboldmath
         Production Limits}
\label{limits}

The limit is calculated assuming that the candidate sample in the
signal region is composed of a large background and a small signal
sample. Each contribution has a Poisson probability density function
although the background can be approximated by a Gaussian distribution
${\cal G}(n; e_{\mathrm{bck}}, \sigma)$ since the background
expectation is large. The width of the Gaussian, $\sigma = 28$, has
two contributions, one accounting for the statistical uncertainty in
the signal window, the other one taking into account the fit error.
$n_0 = 610$ candidates are observed in the signal window. The
probability to count $n_0$ candidates or less is given by
\begin{equation}
  P(n_0; e_{\mathrm{sig}}) \ = \ \sum\limits^{n_0}_{n=0} 
           \ \sum\limits^n_{n_{\mathrm{sig}}=0} 
           \ {\cal P}(n_{\mathrm{sig}}; e_{\mathrm{sig}})
           \ {\cal G}(n-n_{\mathrm{sig}}; e_{\mathrm{bck}}, \sigma)
\end{equation}
where $e_{\mathrm{sig}}$ is the unknown expectation value of the
signal and ${\cal P}$ its Poisson distribution.  The summation is
performed over all possible combinations of $n$, the total number of
candidates in the mass window, and $n_{\mathrm{sig}}$, the amount of
signal within these $n$ candidates, where $n\leq n_0$.  Assuming no
narrow $\Dsprimep$ are produced, i.e.~$e_{\mathrm{sig}} = 0$, a
probability of $0.70$ is obtained to observe 610 candidates or less.
Given the prior knowledge of the maximum amount of background, the
95\%~C.L.~limit is obtained at $P(n_0; e_{\mathrm{sig}}) =
0.70\times0.050$. At large $n$ ($\approx 10$) the systematic
uncertainties affecting the efficiency start playing a role in
calculating the limit. They are considered by substituting the Poisson
distribution of the signal with a Gaussian distribution at $n>10$ and
adding in quadrature the asymmetric systematic uncertainty of the
efficiency to the width of the Gaussian given by $\sqrt{n}$. The
obtained limit on the production of narrow $\Dsprimepm(2629)$ in
hadronic $\Z$ decays is:
\begin{equation}
    f(\Z \to  \Dsprimepm(2629)) \times Br(\Dsprimep \to \Dstarp \pionp
    \pionm) < 3.1 \times 10^{-3} \ \ (95\%\ \mathrm{C.L.}).
\end{equation}
This corresponds to a $95\%$~C.L. upper limit of 66 on the number of
reconstructed $\Dsprimep$ in the signal region. The approach used here
is similar to a Bayesian approach with a flat prior distribution for
positive $\Dsprimep$ production rates and zero elsewhere, but superior
in the fact that negative expectation values are excluded in principle
by using the proper Poisson distribution.

Limits on the $\Dsprimep(2629)$ production rate in charm and bottom
events are also calculated. In the charm and bottom enriched samples,
the numbers of events in the signal region, relative to the
expectations derived from the fitted background functions, are
$5\pm18$ and $29\pm23$, respectively. In using these results to
calculate limits, a further systematic uncertainty arises due to the
experimental uncertainties on \mbox{$\GGcc=0.1671\pm0.0048$} and
\mbox{$\GGbb=0.21644\pm0.00075$}~\cite{pdg}. Under the conservative
assumption that $\Dsprimep$ are only produced in $\Z\to\cc$ events,
i.e. that the measured excess can be fully assigned to $\Dsprimep$
candidates in $\cc$ events, a production limit of
\begin{equation}
    f(\qc \to  \Dsprimep(2629)) \times Br(\Dsprimep \to \Dstarp \pionp
    \pionm) < 0.9 \times 10^{-2} \ \ (95\%\ \mathrm{C.L.})
\end{equation}
is obtained from the charm enriched sample, while under
the opposite assumption that all $\Dsprimep$ are only produced in the
decay of $\qb$-hadrons, a limit of
\begin{equation}
    f(\qb \to  \Dsprimep(2629)) \times Br(\Dsprimep \to \Dstarp \pionp
    \pionm) < 2.4 \times 10^{-2} \ \ (95\%\ \mathrm{C.L.})
\end{equation}
is computed using the bottom enriched sample. The systematic
uncertainties assumed for the calculations in the separate samples are
given in Table \ref{systerr}.


\section{Discussion}

\begin{figure} [htb]
  \center{ \epsfig{file=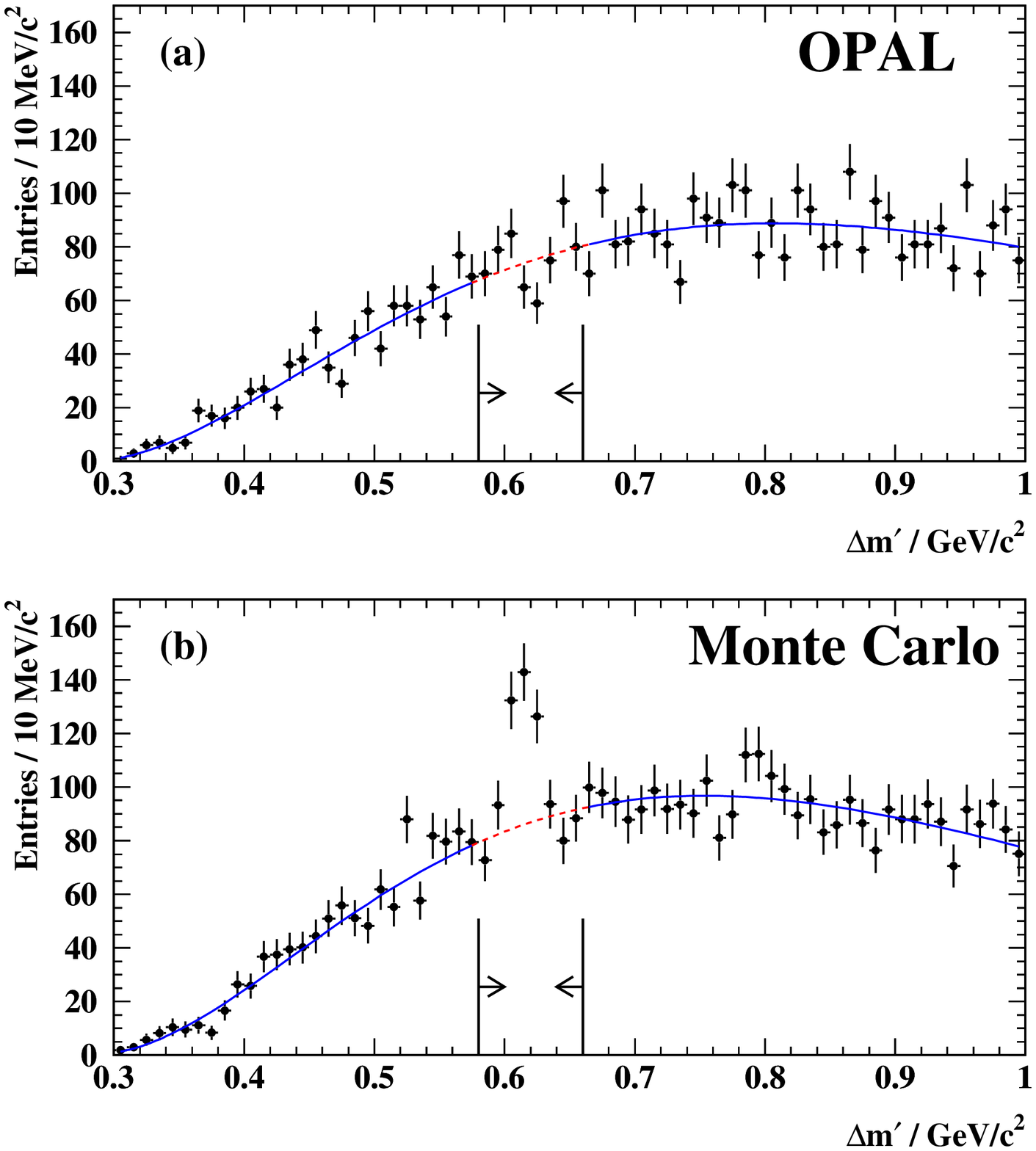,height=19cm}
    \smcap{\label{mpdistr} Distribution of the mass difference
      $\dmp=\mDsp-\mDstar$ of $\Dsprimep$ candidates in the combined
      charm and bottom sample, for (a) data and (b) Monte Carlo. The
      calculation of $\dmp$ as well as the background parametrisation
      superimposed on the histogram are described in the text. The
      Monte Carlo histogram is scaled to the number of hadronic events
      in data. Furthermore, the $\Dstarp$ Monte Carlo production rate
      has been adjusted to an OPAL measurement of this quantity
      \cite{rcpaper}, and the $\Dsprimep$ production rate has been
      fixed at the value published by DELPHI \cite{delphidsp} and the
      OPAL $\Dsszero$ measurement \cite{opaldss}.  The Monte Carlo
      plot presented here is created from a subsample of the available
      events that has roughly the same size as the data sample. The
      arrows indicate the mass window defined as the signal region.
      The line represents the result of the background fit, where the
      line is dashed in regions that have been excluded from the fit.
      The $\chi^2$/dof of the background fit is 1.23 for data, and
      0.99 for the Monte Carlo simulation.}  }
\end{figure}

The result of the OPAL $\Dsprimep$ search does not agree with the
published DELPHI results \cite{delphidsp}.  Both analyses apply
standard selection criteria to obtain a high-purity $\Dstarp$ sample.
In contrast to the OPAL analysis, DELPHI reconstructs $\Dzero$
candidates in two decay channels, $\Dzero\to\Km\pionp$ and
$\Dzero\to\Km\pionp\pionp\pionm$.  The loss in efficiency associated
with using only the $\Km\pionp$ channel is in part compensated by the
softer kinematic cuts and a slightly larger hadronic event sample.
Combinatorial background is reduced by applying a best candidate
selection to $\Dstarp$ candidates which share tracks. These candidates
can produce correlated entries in the $\dmp$ distribution if they are
combined with the same $\Dsprimep$ pions. The expected sensitivity of
the analysis to the existence of a narrow resonance in the mass region
including the theoretical predictions for the $\Dsprime$ as well as
the mass of the published DELPHI result, produced at a rate published
by DELPHI, is demonstrated in Figure~\ref{mpdistr}.  The excess
observed in the Monte Carlo sample with comparable size to the data
sample has a significance of $4.7\stddev$ (statistical error only), or
$3.5\stddev$ including all systematic uncertainties.

The OPAL and DELPHI selection criteria used to select charm and bottom
enriched samples, and for the pions from the $\Dsprimep$ decay, have
been optimized in different ways.  For example, the kinematic cuts for
the pion candidates from the $\Dsprimep$ are less restrictive,
$0.5\GeVc$ in this analysis while DELPHI requires $1.0\GeVc$.

The sensitivity of the OPAL analysis has been checked with the
$\Dsszero$ reconstruction described above, where agreement between
data and simulation is observed.  The selection criteria that are
specific to either the $\Dsprimep$ or $\Dsszero$ analysis are checked
by performing the analysis without these cuts.  This did not change
the result of the $\Dsprimep$ search.

Overall, the sensitivities of the two analyses are found to be
similar, although the background in this analysis is higher.  A
comparison between the analyses can be made by calculating the
$\Dsprimep$ versus $\Dsszero$ rate for which DELPHI published a
number~\cite{delphidsp}:
\begin{eqnarray}
    R&=&
    {{\langle\mathit N_{\Dsprimep}\rangle \mathrm
    Br(\Dsprimep\rightarrow\Dstarp\pionp\pionm)} \over {\langle\mathit
    N_{\Dzeroone}\rangle\mathrm Br(\Dzeroone\rightarrow\Dstarp\pionm)+
    \langle\mathit N_{\Dstwozero}\rangle\mathrm
    Br(\Dstwozero\rightarrow\Dstarp\pionm)}} \\
    \nonumber
     &=& 0.49 \pm 0.18
    \mathrm{(stat.)} \pm 0.10 \mathrm{(syst.)}
    \ ,
\end{eqnarray}
where $\langle N_{\Dsprimep}\rangle$ denotes the expected number of
$\Dsprimep$ in a sample with $\langle N_{\Dzeroone}\rangle$ and
$\langle N_{\Dstwozero}\rangle$ being the corresponding quantities for
the $\Dzeroone$ and $\Dstwozero$. Based on the numbers presented in
Sections \ref{results} and \ref{systematics},
\begin{equation}
  R=0.05\pm0.10\mathrm{(stat.)}\pm0.002\mathrm{(syst.)}
\end{equation}
is calculated. The statistical error of the fit and the systematic
uncertainties resulting from the $\Dsprime$ width and relative
production rates in b and c are included. In the ratio calculation,
all other systematic uncertainties largely cancel and have therefore
been neglected. Using a statistical approach analogous to the one
described earlier, a limit of
\begin{equation}
  R<0.22 \ \ (95\%\ \mathrm{C.L.})
\end{equation}
is computed. 

In summary, the evidence of $\Dsprimep$ production in hadronic decays
of the $\Z$ published by the \mbox{DELPHI} collaboration
\cite{delphidsp} is not confirmed with OPAL data.

\appendix

\ \\
\noindent
{\Large \bf Acknowledgements}

\noindent
We particularly wish to thank the SL Division for the efficient operation
of the LEP accelerator at all energies
 and for their continuing close cooperation with
our experimental group.  We thank our colleagues from CEA, DAPNIA/SPP,
CE-Saclay for their efforts over the years on the time-of-flight and trigger
systems which we continue to use.  In addition to the support staff at our own
institutions we are pleased to acknowledge the  \\
Department of Energy, USA, \\
National Science Foundation, USA, \\
Particle Physics and Astronomy Research Council, UK, \\
Natural Sciences and Engineering Research Council, Canada, \\
Israel Science Foundation, administered by the Israel
Academy of Science and Humanities, \\
Minerva Gesellschaft, \\
Benoziyo Center for High Energy Physics,\\
Japanese Ministry of Education, Science and Culture (the
Monbusho) and a grant under the Monbusho International
Science Research Program,\\
Japanese Society for the Promotion of Science (JSPS),\\
German Israeli Bi-national Science Foundation (GIF), \\
Bundesministerium f\"ur Bildung und Forschung, Germany, \\
National Research Council of Canada, \\
Research Corporation, USA,\\
Hungarian Foundation for Scientific Research, OTKA T-029328, 
T023793 and OTKA F-023259.\\



\clearpage

\end{document}